# Learning From Alarms: A Robust Learning Approach for Accurate Photoplethysmography-Based Atrial Fibrillation Detection using Eight Million Samples Labeled with Imprecise Arrhythmia Alarms

Cheng Ding[1,4], ZhiCheng Guo[2], Cynthia Rudin[3], Ran Xiao[4], Amit Shah[5,6], Duc H. Do[7], Randall J Lee[8], Gari Clifford[1,9], Fadi B Nahab[10], Xiao Hu[1,4,9]

[1]Department of Biomedical Engineering, Georgia Institute of Technology, Atlanta, GA, USA
[2]Department of Electrical and Computer Engineering, Duke University, Durham, NC, USA
[3]Department of Computer Science, Duke University, Durham, NC, USA
[4]School of Nursing, Emory University, Atlanta, GA, USA
[5]Dept of Epidemiology RSPH, Emory University School of Medicine, Atlanta, GA, USA
[6]Department of Medicine, Emory University School of Medicine, Atlanta, GA, USA
[7]School of Medicine, University of California at Los Angeles, Los Angeles, CA, USA
[8]School of Medicine, University of California at San Francisco, San Francisco, CA, USA
[9]Department of Biomedical Informatics, Emory University School of Medicine, Atlanta, GA, USA
[10]Department of Neurology, Emory University School of Medicine, Atlanta, GA, USA

## Abstract

Atrial fibrillation (AF) is a common cardiac arrhythmia with serious health consequences if not detected and treated early. Detecting AF using wearable devices with photoplethysmography (PPG) sensors and deep neural networks has demonstrated some success using proprietary algorithms in commercial solutions. However, further advancement of this paradigm of continuous AF detection in ambulatory settings, towards a population-wide screening use case, still faces several challenges, one of which is the lack of large-scale labeled training data. To address this challenge, in this study, we propose to leverage AF alarms from bedside patient monitors to label concurrent PPG signals, resulting in the largest PPG-AF dataset so far (8.5M 30-second records from 24100 patients) and demonstrating a practical approach to build large labeled PPG datasets. Furthermore, we recognize that the AF labels thus obtained contain errors because of false AF alarms generated from imperfect built-in algorithms from bedside monitors. Dealing with label noise with unknown distribution characteristics in this case requires advanced algorithms. We, therefore, introduce and open source a novel loss design, the cluster membership consistency (CMC) loss, to mitigate label errors. By comparing CMC with state-of-the-art methods selected from a noisy label competition, we demonstrate its superiority in multiple aspects including handling label noise in PPG data, resilience to poor-quality signals, and computational efficiency.

## Introduction

Atrial fibrillation (AF) is the most common cardiac arrhythmia, with an estimated prevalence of 1-2% in the general and 9% in the aging population (age > 65 years), respectively [1]. Even though AF can be asymptomatic and transient (i.e., paroxysmal AF -- pAF), it can implicate future serious health conditions such as ischemic stroke, congestive heart failure, and other high-risk cardiopulmonary conditions 0.941. Therefore, there is a growing need for industry and academia to develop solutions that can detect pAF in free-living (ambulatory) settings [3]. Among signal modalities to detect pAF, photoplethysmography (PPG) is well-suited for passive monitoring due to its strong physiological premise and the practical consideration that PPG sensors are ubiquitously available in more than 71% of consumer wearable devices [4]. The physiological premise is that irregular heartbeats generate variable cardiac output so that peripheral pulsatile blood volume changes follow the same pattern, resulting in irregular pulse-to-pulse intervals and irregular pulse morphologies in PPG.

Deep neural networks (DNNs) have been widely applied in many healthcare domains including PPG-based AF detection [5-13]. They are currently the most promising approach to achieve accurate PPG-based AF detection. Although the problem and feature distributions may differ, it remains valuable to explore the potential benefits of training DNN-based PPG-AF algorithms on datasets that are comparable in size and diversity to those in other domains (i.e., computer vision). However, building a dataset with large sample size and rich diversity at both PPG signal and patient levels is a challenging task. One study [12] used 500,000 30-second PPG strips, but these were from only 81 patients. While the dataset used in another study [8] was based on 3373 patients, the total number of PPG strips was 186,317, i.e., 15 minutes of PPG recordings from each patient, which is relatively limited amount of data available per individual, potentially impacting the algorithm's ability to capture individual-specific variations in PPG-AF patterns. The rest of the studies [5, 9, 12, 13] used even fewer PPG strips and patients. In addition to this issue, arrhythmias like premature atrial contraction (PAC) and premature ventricular contraction (PVC) often confuse AF detection algorithms because they cause irregular pulse-to-pulse intervals similar to those in AF. Therefore, it is important to include more PPG signals, also covering those with AF-mimicking arrhythmias to mitigate false positive detection in the real-world setting [14]. Existing studies have not addressed this issue, and many of them tested their algorithms on data from only laboratory settings rather than ambulatory settings [5, 7, 8, 13, 15-20].

In order to increase training data size, many groups have used data augmentation techniques [21], such as generative adversarial networks (GANs), to augment training data. In particular, previous work developed a GAN with a novel generator architecture and a novel power spectrum-based loss function for PPG-based AF detection [22]. We observed that augmented training data was able to increase accuracy, but performance quickly saturated once the number of augmented samples reached 2 million. In other words, augmented data was not as helpful as more real data might have been. **Therefore, the present study continues to address this training data issue by introducing two innovations: (1) leveraging PPG signals collected from bedside patient monitors and their built-in arrythmia detection algorithms to create a large-scale dataset; and (2) presenting an algorithm that achieves robust learning in the presence of label noise.**

Through the first innovation, we build a large-scale dataset for model development. To do this, we identified PPG strips corresponding to different arrhythmias, as detected by the patient monitor manufacturers' proprietary algorithms. These algorithms tend to be overly sensitive at the cost of specificity. That is, the algorithms will generate alarms when they detect any abnormal ECG patterns, providing noisy labels for the concurrent PPG signals. This approach avoids having to manually annotate millions of records; hence we were able to assemble a training dataset consisting of approximately **8 million 30-second PPG strips from 24100 patients, which is thus far the largest dataset in both sample and patient levels in this field of study**. We focus on AF and various PVC alarms to label PPG data because these two kinds of arrhythmias are the most relevant for training PPG-based AF detectors to be used by asymptomatic patients. We could not find studies that quantify the accuracy of these alarms, and it is reasonable to assume that these alarms are not accurate based on previous studies of other more lethal cardiac arrhythmia alarms [23, 24]. Therefore, as a second innovation, we introduce a novel approach to achieve robust learning in the presence of label noise caused by using inaccurate PPG alarm labels.

Through the second innovation, we offered a robust algorithm that handles label noise. A comprehensive review of previous research studies on label noise can be found in [25] and a more recent curated list of resources can be found at [26], and will be summarized in Section 2.1 below. Here, we would like to point out that we cannot naively apply these existing algorithms in our setting for three reasons. First, most studies do not deal with label noise in real-world situations. The authors conducted the experiments by synthesizing incorrect labels (as label noise) by either symmetrically or asymmetrically flipping the labels. In reality, the distribution of real-world label noise is much more complex, and thus does not satisfy many of the assumptions of these algorithms. In the proposed dataset, factors such as changes in either monitoring hardware or the built-in alarm algorithms could affect labeling consistency, leading to varying levels of label noise which cannot be easily estimated. Second, few studies have considered the relationship between data quality (i.e., signal quality in our case) and label noise. We were only able to find one relevant work [27], in which authors tried to partition the input features into several uncorrelated groups and reduce the weight of the less informative ones based on their mutual information with the noisy labels. This approach leverages the natural clustering of the data, a key factor in accurately identifying the label noise transition matrix. Low-quality signals are ubiquitous when collecting PPG signals due to patient movement or loose contact of sensors from the device, which makes signal quality an inevitable barrier for PPG signal processing. Third, sophisticated algorithms may lead to higher performance but are hard to scale up. For instance, Co-teaching [28] and DivideMix [29] train two models simultaneously, which doubles the training cost. To address the above three concerns, we proposed a simple but effective loss function: cluster membership consistency (CMC) loss. We compare prediction performance with the state-of-the-art methods, including Co-teaching [28], Early-Learning Regularization (ELR) [31], DivideMix [29], Iterative Noisy Cross-Validation (INCV), and Sparse Over-Parameterization (SOP) [32], which are top runners in the 1st Learning and Mining with Noisy Labels Challenge at [33]. The performance of our proposed methods are evaluated through comprehensive experiments on three PPG datasets from external institutions.

## 2 Related work

### 2.1 Related work on Algorithms for Reducing Label Noise

**Neural network architecture design for robustness.** Several studies have used dedicated model designs to achieve robust learning [28, 29, 34]. These methods usually use one of two types of networks. Probabilistic noise modeling [34] attempts to model the label transition probability. Co-teaching [28] trains networks with two branches. Each branch selects low-loss samples for the other branch for further training. For instance, DivideMix [29] leverages two branches of networks. Each branch divides the data into clean and noisy portions for the other branch and adopts a semi-supervised training schema. These methods tend to be computationally expensive with dual-network training at the same time, which are hard to scale when the size of training sample getting larger.

**Regularization for robustness.** Regularization has been a crucial technique in combatting noisy labels and overfitting for deep learning models. Recent work developed the Mixup algorithm [35], which uses linearly-interpolated samples as additional training samples. Deep bilevel learning [36] leverages a clean validation set and sets weights for each mini-batch to optimize over the clean validation set to reduce overfitting. Both [37, 38] regularize models against the effect of memorization on the validation set. Authors of Robust Early-Learning [38] discover that only a portion of the model parameters are crucial for generalization and learning on clean labels and other parameters are more likely to fit noisy labels. Thus, the model parameters were divided into critical and non-critical sets and these two sets of parameters were updated differently. ELR [37] leverages "early learning" to alleviate the effect of noisy labels later in the training process; they assume that the important information is learned first and stops before the model starts fitting the noise. Open-Set Samples with Dynamic Noisy Labels (ODNL) [39] artificially creates hard samples at each epoch by assigning randomly selected labels to an auxiliary dataset that contains both seen labels in the noisy label set and previously unseen clean labels, thus preventing the deep model from overfitting noise samples.

**Sample selection for robustness.** Sample selection is an intuitive way to solve the label noise problem. It aims to distinguish between correctly labeled and incorrectly labeled samples. Sample selection approaches leverage the observation that before the model overfits, high-loss samples are more likely to be noisy samples and low-loss ones are likely to be clean samples. INCV [40] attempts to select correctly labeled samples based on validation loss values. At each iteration of the INCV algorithm, it randomly splits the dataset into training and validation partitions, and high loss samples in the validation partition are discarded as noisy samples. By iterative removal of high-loss samples (noisy samples), it will produce a set of clean samples for training. Several works [28, 41] also rely on loss values to select clean samples. Even though sample selection can achieve good results in ideal situations, in practice, mistakes in the selection process accumulate and amplify the error in the end.

**Loss functions for robustness.** Popular loss functions, such as cross-entropy loss, perform well in noise-free settings, however given noisy labels, these popular loss functions may not be the optimal choice. Modifying or improving loss functions has shown to be effective in combating noisy labels. GCE loss [39] combines the robust nature of mean absolute error (MAE) and the advantage of traditional cross-entropy loss. Recently, SOP from Liu et al. [32] operates based on the assumption that a network learned on a clean label set will not predict well given label noise when the label noise is likely to be sparse. SOP aims to model the label noise and learn to separate the noisy data from the clean data. In particular, SOP modified the classic cross-entropy loss with an additional sparseness term to model the label noise of each sample. Additional sparseness term to model the label noise of each sample based on the sparse nature of the label noise. Symmetric Cross Entropy (SCE) [42] introduces a reverse CE term to the loss function, (i.e., given two distributions p and q, CE=H(q, p) and RCE=H(p, q) where H() denotes cross-entropy). It improves learning on hard samples and robustness towards label noise. Curriculum loss (CL) [43] provides a tight upper bound for robust 0-1 loss, as well as an extension for multi-class label noise. Robust loss function designs can readily work with existing network designs.

## 2.2 Robust learning on physiological signals

In addition to the above-mentioned studies, several works focused on the label noise problem in physiological signals. Vázquez et al. [44] used self-learning for label correction for multiple-lead ECG signals to reduce the effect of label noise. This approach learns to correct noisy labels during training by comparing samples against a set of randomly chosen prototypes. This process is less than optimal; since the features used to select and compare the prototypes are learned from the noisy labels, the label correction result may be wrong. The performance of this method relies largely on the knowledge of the percentage of noise in the dataset, which is also not applicable in our task. The study [45] also considered cluster information to handle label noise; the authors used K-nearest neighbors (KNN) iteratively to determine if a label of a sample is correct. Due to the nature of KNN and the iterative design of the algorithm, it is not scalable, thus making it unsuitable for our large-scale dataset without reducing the data dimension first. Another study [46] considered an ensemble approach to combine the predictions from a set of under-trained models for sleep apnea using PPG. They share the same assumption that the early learning stage is less likely to be influenced by label noise. We do not compare to this specific method, but we compare against more generalized methods such as the previously mentioned ELR [31]. More recently, Negative-ResNet [47] leveraged a semi-supervised negative learning scheme on noisy-labeled ECG signals. This approach learns from negative labels (i.e., labels other than the true label for each sample) instead of positive labels (i.e., the true label for each sample). The authors assumed in multiclass noisy label settings negative labels are more likely to be correct than positive labels. This approach may be an excellent choice for multiclass label noisy learning; however, the multiclass negative learning assumption in this study does not hold for our binary classification task.

# 3 Study data

The study data used for this study comprised a combination of retrospective private data and publicly available data. The data for model development was collected from routine use of bedside patient monitors in intensive care units at the University of California San Francisco (UCSF) medical center (Institution A) with an approved waiver of written patient consents by the UCSF institutional review board (IRB number:14-13262). The retrospective use of the fully de-identified version of this data archival in the current project is conducted according to the terms of a data use agreement between UCSF and Emory University and hence the dataset cannot be made public without additional institutional approvals.

Three datasets from external institutions were used for model testing. The first test dataset was collected also from routine use of bedside patient monitors in acute care units at the University of California Los Angeles medical center (Institution B) with a waiver of written patient consents approved by the IRB (10-000545). The second one was collected from patients undergoing AF ablation procedures at the Emory University Hospital who consented to wear a study device to obtain PPG signals under IRB (00084629). The retrospective use of the de-identified versions of these two datasets has been approved by the corresponding institutional IRBs. The third dataset was from a public data repository [48]. Detailed information for each dataset is reported in the following sections.

## 3.1 Study population from Institution A – Model development

A waveform dataset, including physiological signals, cardiac arrhythmia alarms, and linked electronic health records (EHR) from over 24,100 patients were collected at Institution A between 2013 – 2018. The demographics of patients in this dataset are reported in Table 1.

AF alarms were found among 2834 patients in the waveform dataset, out of which 1650 had a documented ICD-10 code for AF (I48) from the linked EHR dataset. In this study, we focus on AF and premature ventricular contraction (PVC) alarms to sample and label the large training dataset. A detailed definition of the alarms is summarized in Table 2.

Table 1: Characteristics of Participants Enrolled in Institution A

| Characteristics | Total Cohort (**N = 28539**) | AF Group (**N = 5779**) |
|---|---|---|
| Sex | | |
| Female | 13203 (46.2%) | 2304 (39.9%) |
| Male | 15330 (53.7%) | 3473 (60.1%) |
| Others | 6 (0%) | 2 (0%) |
| Age | | |
| ≥65 yr | 12157 (42.6%) | 4251 (73.6%) |
| 55-64 yr | 5370 (18.8%) | 928 (16.1%) |
| 40-54 yr | 4372 (25.3%) | 419 (7.3%) |
| 22-39 yr | 2715 (9.5%) | 147 (2.6%) |
| <22 yr | 3925 (13.8%) | 34 (0.6%) |
| Race or ethnic group | | |
| White or Caucasian | 15890 (55.7%) | 3311 (56.5%) |
| Black or African American | 2159 (7.4%) | 435 (7.4%) |
| Asian | 4364 (15%) | 1097 (18.7%) |
| Unknown/Declined | 1149 (4%) | 191 (3.3%) |
| Others | 4913 (16.9%) | 671 (11.4%) |
| Native Hawaiian or Other Pacific Islander | 426 (1.46%) | 125 (3.1%) |
| American Indian or Alaska Native | 212 (0.7%) | 35 (0.6%) |

Table 2: Schema for arrhythmia alarms.

| Alarms | Short label | Definition |
|---|---|---|
| Atrial Fibrillation | **AF** | Irregular timing of QRS complexes and absence of preceding P waves |
| Normal sinus rhythm | **NSR** | Healthy and regular heartbeat |
| Premature Ventricular Contractions | **PVCs** | **VT>2**: 3-5 consecutive ventricular beats at a rate≥100 |
| | | PVC: Isolated PVCs |
| | | **R on T**: PVC falls on the ST or T wave portion of the previous beat |
| | | **Couplet**: Two consecutive PVCs with a rate ≥ 100 |
| | | **Bigeminy**: PVC alternates with a non-ventricular beat for ≥ 3 cycles |
| | | **Trigeminy**: PVC alternates with 2 non-ventricular beats for ≥ 3 cycles |
| | | **PVC≥X**: PVC count is equal to or greater than the user-defined limit |

### 3.1.1 Labeling process

We store 30 seconds of each PPG signal centered at the start time of an AF alarm, as illustrated in Fig. 1; PVC samples are extracted similarly, except that we exclude PVC alarms that have AF alarms within 30 seconds of their onsets, so we do not have samples with multiple labels. Since there is no alarm for normal sinus rhythm, we say that one 'Normal sinus rhythm (NSR) alarm' exists if the time gap between two consecutive cardiac arrhythmia alarms from patient monitors is longer than 30 seconds. With this strategy, we obtain 4,070,350 30-second PPG segments from 2985 patients for AF. For non-AF samples, we randomly selected 2,436,460 NSR samples and 2,157,599 PVC samples from 24,100 patients (some of whom had AF alarms as well). Therefore, AF and non-AF samples are approximately matched in size. Note that we never use the same patient in training and test sets – our test data is from different hospitals than the training set.

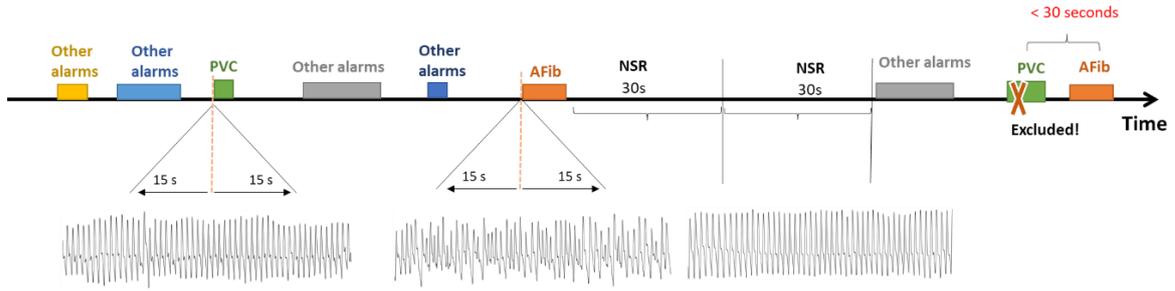

Figure 1: The illustration of the algorithmic labeling process for the three classes. Each box indicates a specific alarm that was triggered over a specific period of time for our example patient. Each time an alarm is triggered, we store 15 seconds of data both before and after the trigger. Note the presence of normal PPG signals (labeled NSR) as well as the presence of many different alarms (VFIB, VT, PVC), and motion artifacts (labeled "artifacts"). On the right, we show that when an AF alarm is triggered within 30 seconds of a PVC alarm, we exclude the PVC alarm, because PVC and AF are difficult to tell apart.

### 3.1.2 Subset of clean-labeled PPG signal from Institution A

Based on our prior study [22], we selected **18,055 PPG** segments from 13 stroke patients from Institution A. Each segment was annotated by at least two clinical experts, with any disagreement adjudicated by another senior cardiologist. This dataset will be used in an experiment where we train with a small, accurately labeled dataset. It should be noted that these 13 patients were included in the patient cohort of institution A.

### 3.2 Study population – Model testing (external validation)

Testing is performed on three external datasets (one in the public domain), as summarized in Table 3. We include the percentage of AF and Non-AF, signal type, and the setting of data collection for each dataset. For data collected from Institution B and data collected from Emory University, the percentage of good quality signals is calculated from predicted labels of a previously developed binary PPG quality assessment model [49]. For Test set 3 (Stanford dataset), the percentage of good quality signals is calculated from the signal quality label provided by the authors.

**Test set 1 - Institution B waveform dataset:** This dataset contains continuous PPG signals collected via pulse oximeter from 126 hospitalized patients, which includes 41 female (32.5%) and 85 male (67.5%). Among all the patients. 80 patients (63.5%) are reported White. Asian, Hispanic and Black patients are 6 (4.7%), 22(17.4%) and 12 (9.5%), respectively. The race of remaining patients is recorded as 'Others'. These patients were admitted between April 2010 and March 2013, with ages ranging from 18 to 95 years old. All model testing procedures were carried out by the co-author affiliated with Institution B. For each continuous recording, the start and end time of each AF episode were annotated by this co-author who is a board-certified cardiac electrophysiologist.

**Test set 2 - Simband dataset from Emory University.** This dataset was used in a prior study [50] where 115 patients were recruited. 72 patients (62%) are male, and 43 patients (38%) are female. 83 patients (72%) are reported as Black or African American, and 26 (22%) are reported as White. The race of remaining patients is not reported. These patients were admitted between October 2015 and March 2016, aged between 18-89 years old. Approximately 5 minutes of PPG was collected from each patient using Samsung Simbands in a hospital setting.

**Test set 3 - Stanford dataset:** This publicly available dataset was described in [48]. The authors recruited 107 patients diagnosed with AF (mean age 68), and 15 more patients (mean age 67) with paroxysmal AF. Data from 41 healthy participants (mean age 56) undergoing exercise stress tests were also included.

Table 3: Characteristics of three external test sets

|  | Test set 1 | Test set 2 | Test set 3 |
|---|---|---|---|
| #Patients | 126 | 115 | 173 |
| #PPG-AF | 38,910 (15%) | 348 (40.7%) | 52911 (39.6%) |
| #PPG-NonAF | 220,740 (85%) | 506 (59.2%) | 80620 (60.4%) |
| good quality(%) | 58.5% | 20.1% | 20.3% |
| Signal type | Fingertip | Wrist-worn | Wrist-worn |
| Setting | Hospital | Hospital | Ambulatory |

# 4 Methods

## 4.1 Problem definition

We denote the training data as $D = \{(x_i, y_i)\}_{i=1}^{|D|}$, in which $x_i$ represents the $i^{th}$ signal and $y_i$ is its label. Here, $y_i \in \{0,1\}$ is a categorical label (0: non-AF and 1: AF). Given the presence of label noise, $y_i$ can be incorrect. Moreover, the distribution of label errors is unknown.

## 4.2 Robust learning with cluster membership consistency

Our proposed algorithm is based on the idea that data with similar latent features extracted by an (*unsupervised*) autoencoder should have similar labels in a *supervised* neural network, as demonstrated in Fig. 2. This is essentially a smoothness assumption on signals that are close on the manifold of realistic signals. An autoencoder is used to map the manifold of real signals to a space where distances along the manifold become Euclidean distances. We then use a clustering approach to group samples into a finite number of clusters.

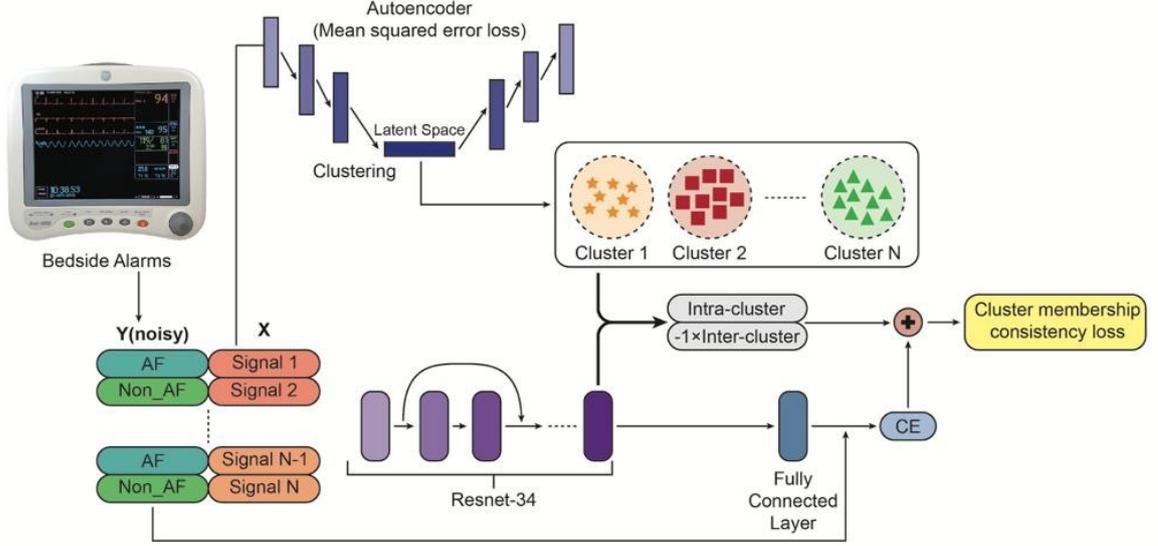

Figure 2: The overall flow of the cluster membership consistency loss. A pre-trained cluster label is assigned to each sample. Based on the cluster labels, intra-cluster distance and inter-cluster distances of each batch are calculated and used as additional loss terms together with conventional cross-entropy.

This unsupervised approach (an autoencoder and clustering) uses only the PPG signals, and not the labels, so it is not affected by noise in the labels. The autoencoder compresses the signals to encode the information into a space where Euclidean distances can serve as meaningful distances along the manifold of realistic signals. The learned cluster membership is used in the subsequent supervised learning phase to encourage distances between points in the same cluster to be small and distances between points in different clusters to be large. This is accomplished by augmenting the cross-entropy loss with two terms, which comprise the cluster membership consistency loss:

$$L_{intra} = \sum_{c=1}^{K}(\sum_{i=1}^{N_c}\sum_{j\neq i}^{N_c}\|F(x_i^c;\theta) - F(x_j^c;\theta)\|) \quad (2)$$

$$L_{inter} = -\sum_{c_1=1}^{K}\sum_{c_2\neq c_1}^{K}(\sum_{i=1}^{N_{c_1}}\sum_{j=1}^{N_{c_2}}\|F(x_i^{c_1};\theta) - F(x_j^{c_2};\theta)\|) \quad (3)$$

where the first term is the sum of distances among intra-cluster samples and the second term is the sum of distances among inter-cluster samples, where the superscript $c$ represents the cluster membership. Distances are computed in a latent space defined by a function $F$ that is parameterized by $\theta$. This distance is easy to compute for any neural architecture and requires little extra computational cost for each training epoch. The final loss is a weighted combination of the cross-entropy loss and two new loss terms,

$$L = L_{CE} + \lambda_1 L_{intra} + \lambda_2 L_{inter}, \quad (4)$$

$\lambda_1$ and $\lambda_2$ are hyperparameters selected by grid search through cross-validation. The detailed training procedure is reported below.

---

**Algorithm 1 Training procedure with cluster membership consistency loss**

**Input**: $\{x_i, y_i\}$ Dataset; **M**, number of clusters; $\varphi$, model parameter set for Resnet; $\lambda_1$ and $\lambda_2$ loss function parameters.

1. Generate embedding $\{h_i\}$ for each $x_i$ by an autoencoder
2. Perform **M**-cluster K-means on $\{h_i\}$, and then generate $\{\hat{y}_i\} \in \{1, \ldots, M\}$ for each $x_i$.
3.     **while** not converged **do**
4.         Calculate $L_{intra}, L_{inter}$ for each cluster in $\{1, \ldots, M\}$
5.         Loss = $L_{CE} + \lambda_1 L_{intra} + \lambda_2 L_{inter}$
6.         Update $\varphi$ through backpropagation
7.     **end while**

---

## 4.3 Data preprocessing and training details

Continuous PPG recordings from the Institution B waveform dataset and Simband data were divided into consecutive 30-second segments without overlap. Since signals from the Stanford dataset are already truncated into 25-second intervals, we augmented each signal by appending the first 5 seconds to the end. Since PPG signals collected from Institutions A and B were both sampled with 240 Hz, we upsampled data from test sets 2 and 3 to 240 Hz. Min-max normalization was then applied to scale each segment to be within [0,1].

We used Resnet-34 as the base model for all methods, including CMC and other baseline models. We trained each network for 50 epochs with a batch size of 1000 samples. The validation set was generated by using samples from 20% of the patients randomly selected from the Institution A dataset; the remaining 80% of the Institution A data were used as the training set. The best epoch was selected as that with the least loss on the validation set and the corresponding model was retained. Adam was used as the optimizer with a learning rate of $10^{-5}$. INCV, SCE, ELR, and SOP were trained on the GPU cluster from the Emory Biomedical Informatics department with multiple GPUs. Since it is memory intensive, INCV was trained on a large memory machine with an Nvidia Tesla P100, and SCE, ELR and SOP were trained with an Nvidia Tesla V100. Other methods were trained on a standalone server with one Intel Xeno 6212U (48 cores) and one Nvidia A2 Tensor Core GPU.

## 5    Experiments and Results

### 5.1  Large-scale data with noisy labels vs. small-scale data with clean labels

In this experiment, we want to demonstrate the value of large-scale datasets, even though they contain label noise. We compared models trained with large-scale data with label noise to models trained with a small curated data set with clean labels, as described in Sections 3.3 and 3.4. A third dataset was created by augmenting the small dataset with one million synthetic

PPGs based on a validated GAN from our previous study [21]. To gain a holistic understanding of the size of the neural network on classifier performance, we compared classifiers trained with varying training set sizes and model complexities, including Resnet-18, Resnet-34 and Resnet-101 (i.e., an 18-layer, 34-layer, and a 101-layer 1D ResNet). The CE loss was used as the cost function for all models.

As reported in Fig. 3, models trained on the large dataset with noisy labels consistently outperformed models trained with the same algorithms on both the small dataset with clean labels and the dataset augmented with synthetic data from the GAN across different model complexities. Importantly, our experiment indicates that it is beneficial to increase model complexity when more training data are available. For example, on Test Set 2, a large Resnet101 trained on the large noisy dataset achieves an AUROC (Area under the ROC Curve) of 90% vs. an AUROC of 73% obtained using Resnet18. It should be noted that the synthetic data was generated from a GAN proposed in prior work [21] and trained together with a small-scale dataset. Augmenting the training data with synthetic data increases the AUROC but it is less effective than using the large scale data from patients. We observed that synthetic data are not necessarily helpful in training more complex models, e.g., AUROC actually decreases when training more complex models (Resnet 101) compared to simpler models (Resnet-34, Resnet-18) in the presence of synthetic data.

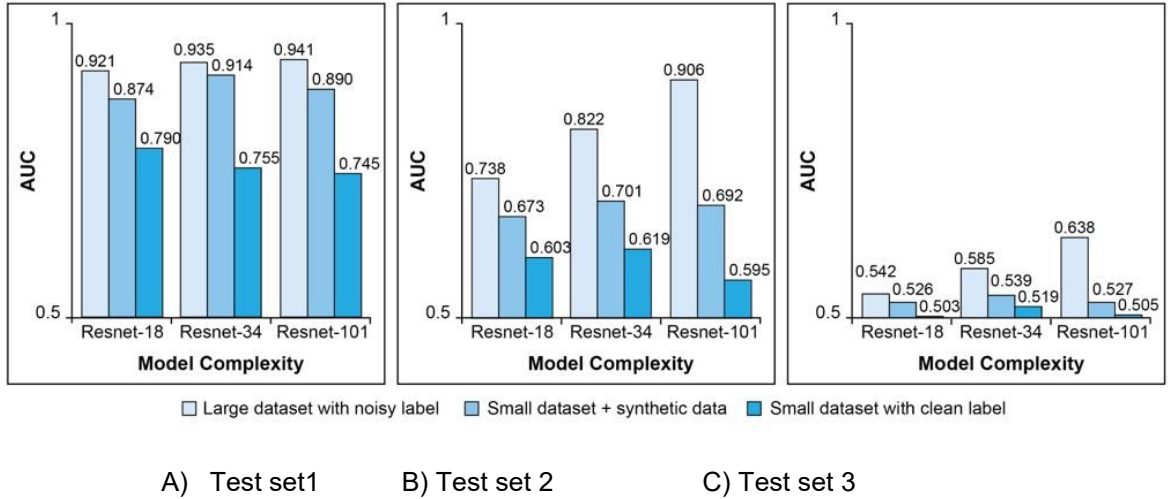

A) Test set1      B) Test set 2      C) Test set 3

**Figure 3.** The experiment on a large noisy dataset vs. a small clean dataset vs. a synthetically augmented clean dataset. Specifically, the datasets were: the large real-world data with noisy labels (8M records), the small scale real-world data with clean labels (1.8K records) and the clean data augmented with synthetic data (1.8K + 1M records). Three different model complexities were also investigated, specifically Resnet-18, Resnet-34 and Resnet-101, respectively.

## 5.2 Performance of the proposed model and its comparison with baseline models.

As the key investigation in this study, we assess the efficacy of the proposed CMC loss against Co-Teaching [28], Early-Learning Regularization (ELR) [31], DivideMix [29], Iterative Noisy Cross-Validation (INCV), and Sparse Over-Parameterization (SOP) [32], which were top performers in the 1st Learning and Mining with Noisy Labels Challenge at [33]. We tested each method on PPG data with both high and poor-quality subgroups, in addition to the entire dataset. We used AUROC as the metric for AF evaluation, the Area Under the Precision-Recall Curve (AUPRC) is reported in Table S1 in the supplementary. Notably, we chose 6 as the cluster number because we used three types of alarms, AF, PVC and NSR, and signals from each alarm type can have good quality and poor quality, which gives us 6 categories total.

Table 4: Performance comparison between the proposed method with state-of-the-art algorithms. Bolded numbers indicate the top performance across all models being compared.

| Method | Test set 1 | Test set 2 | Test set 3 |
|---|---|---|---|
| CE | 0.924 ± 0.02 | 0.836 ± 0.01 | 0.585 ± 0.01 |
| SCE | 0.929 ± 0.02 | 0.843 ± 0.03 | 0.558 ± 0.01 |
| Co-teaching | 0.905 ± 0.01 | 0.824 ± 0.02 | 0.539 ± 0.01 |
| INCV | **0.932 ± 0.01** | 0.861 ± 0.01 | 0.605 ± 0.01 |
| DivideMix | 0.931 ± 0.01 | 0.891 ± 0.01 | **0.737 ± 0.01** |
| ELR | 0.860 ± 0.02 | 0.811 ± 0.01 | 0.566 ± 0.01 |
| SOP | 0.930 ± 0.02 | 0.887 ± 0.02 | 0.661 ± 0.01 |
| CMC (Ours) | **0.932 ± 0.02** | **0.910 ± 0.02** | 0.735 ± 0.01 |

* Tested on the whole dataset

| Method (**AUROC**) | Test set 1 | | Test set 2 | | Test set 3 | |
|---|---|---|---|---|---|---|
| | Good quality | Bad quality | Good quality | Bad quality | Good quality | Bad quality |
| CE | **0.943 ± 0.02** | 0.912 ± 0.02 | 0.934 ± 0.02 | 0.768 ± 0.02 | 0.896 ± 0.02 | 0.542 ± 0.02 |
| SCE | 0.935 ± 0.02 | 0.905 ± 0.02 | 0.929 ± 0.02 | 0.755 ± 0.02 | 0.710 ± 0.02 | 0.531 ± 0.02 |
| Co-teaching | 0.917 ± 0.01 | 0.875 ± 0.01 | 0.905 ± 0.01 | 0.731 ± 0.01 | 0.777 ± 0.01 | 0.500 ± 0.01 |
| INCV | 0.942 ± 0.01 | **0.927 ± 0.01** | 0.942 ± 0.01 | 0.778 ± 0.01 | 0.915 ± 0.01 | 0.547 ± 0.01 |
| DivideMix | **0.943 ± 0.01** | 0.926 ± 0.01 | 0.938 ± 0.01 | 0.819 ± 0.02 | 0.964 ± 0.01 | **0.695 ± 0.01** |
| ELR | 0.869 ± 0.02 | 0.861 ± 0.02 | 0.919 ± 0.02 | 0.684 ± 0.02 | 0.743 ± 0.02 | 0.524 ± 0.02 |
| SOP | 0.937 ± 0.02 | 0.914 ± 0.02 | 0.987 ± 0.02 | 0.797 ± 0.02 | 0.958 ± 0.02 | 0.571 ± 0.02 |
| CMC (Ours) | 0.937 ± 0.02 | **0.927 ± 0.02** | **0.991 ± 0.02** | **0.843 ± 0.01** | **0.974 ± 0.02** | 0.670 ± 0.02 |

* Tested on good and bad quality subgroups

According to the findings presented in Table 4, CMC consistently demonstrates superior or second-best performance across all nine testing scenarios, irrespective of whether we tested on the entire dataset or specific subgroups categorized by signal quality. This conclusion is supported by the bold numbers displayed in the respective columns of the table. Among the various baselines examined, DivideMix exhibits comparable performance and achieves the highest AUROC when tested on Test Set 3. It is worth noting that the disparity in performance values between subgroups with good and poor signal quality is more pronounced in Test Sets 2 and 3, where most signals exhibit bad quality.

## 5.3 The choice of cluster number

In Table 4, we choose 6 as the cluster number for CMC. To evaluate the statistical significance of the choice of cluster number, we employed the Wilcoxon signed-rank test. Using a bootstrapping approach, we randomly selected one sample per patient at each draw, resulting in 126, 173, and 98 test samples for each of the three test sets for calculating the AUROC scores (three test sets each contains 126, 173, and 98 unique patients). We repeated this sampling process 100 times to ensure the robustness of our results. We also include DivideMix since its performance is on par with CMC.

Figure 4 shows the results for the three test sets separately where pairs are color-coded with the corresponding p-values. The threshold for statistical significance (for the p value) is adjusted using the Bonferroni correction from 0.05 to 0.0033 since we conducted multiple pairwise tests (i.e., DivideMix and 5 variants of the CMC method). The CMC-N differences are presented in the lower right-hand corner of each table. It can be observed that there is no significant difference in AUROC between each CMC-N in the three test sets, except that only CMC-6 shows a statistically significant difference in Test Set 1. When comparing with Dividemix, we observe that the cluster numbers do not yield results that are significantly different from each other.

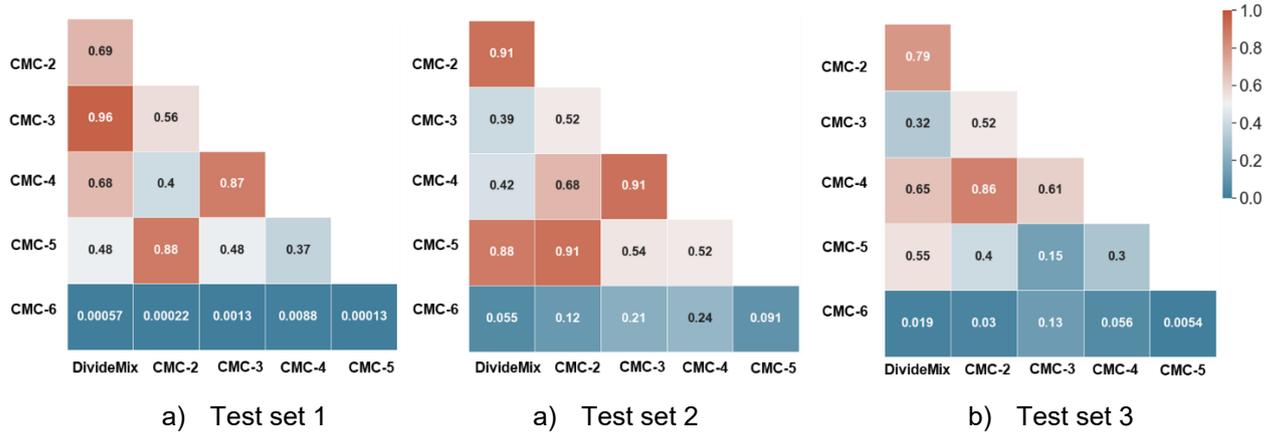

a) Test set 1  a) Test set 2  b) Test set 3

**Figure 4**. P-value of Wilcoxon signed-rank test between AUROC scores of different methods for all datasets. P-values lower than 0.0033 indicate statistical significance in terms of the difference between the tested performance metric.

## 5.3 Robustness to poor quality signals

We investigate the impact of PPG signal quality on performance for each method here. This involves analyzing differences between each method's AUROC scores evaluated on good and bad quality subgroups. These results are derived from the AUROCs obtained in Table 4 and shown in Fig. 5. By exploring the influence of PPG signal quality on the performance of each method, we can gain deeper insights into their ability to withstand changes in signal quality.

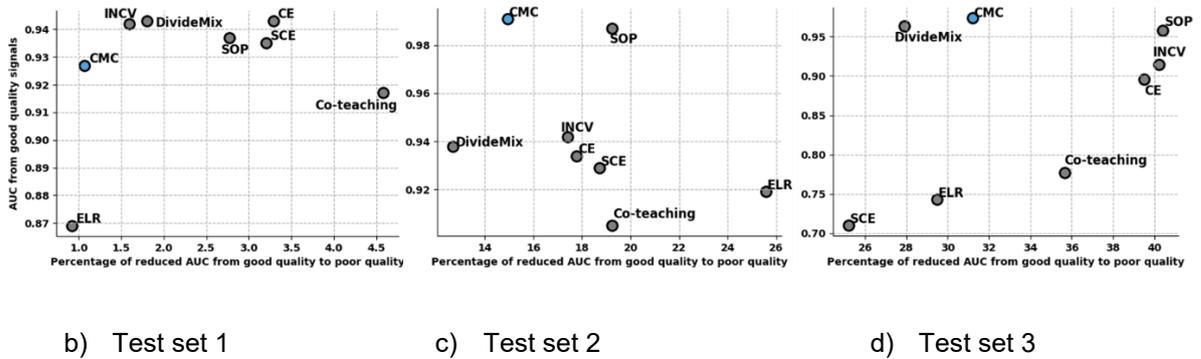

b) Test set 1  c) Test set 2  d) Test set 3

**Figure 5:** The comparison of resilience to poor data quality between methods for each of the three test sets. Data are directly derived from Table 4, the y-axis represents the AUC on good quality subgroups and the x-axis represents the percentage of reduced AUC when test data

switched from good quality to bad quality. The evaluation of a method is considered more favorable when its data points tend to be positioned towards the top-left region of a graph. This positioning indicates that the method achieves a high AUC on good quality signals and simultaneously experiences a minimal reduction in AUC when transitioning to a dataset with poor signal quality.

Results in Table 4 clearly show that signal quality has a large impact on algorithm performance. Figure 5 more directly reveals how PPG signal quality impacts algorithm performance. The X-axis represents the percentage of reduced AUROC when testing on the bad-quality subgroup compared to the good-quality one; the Y-axis represents the AUROC tested on the good quality samples. A good-performing algorithm would provide high AUROC score on good quality signals while offering good resilience to bad signal quality, hence being placed on the top left corner in Fig. 4. All CMC models are near the top-left corner for all three datasets (as coded in blue). ELR has the worst performance, visible at the bottom-right corner for Test Set 2 and Test Set 3, and the bottom left for Test set 1.

## 5.4 Computation time

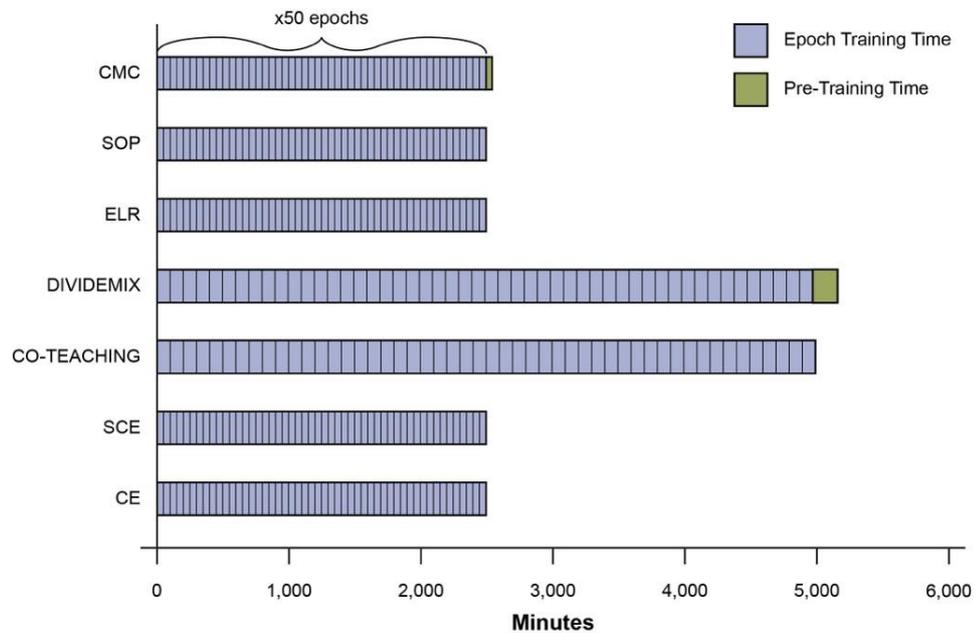

**Figure 6:** Training time of Resnet-34 model for different methods.

Optimizing the performance of deep learning models typically necessitates a substantial amount of training data, and the training time is directly proportional to the size of the dataset. Therefore, the computational efficiency of the algorithms becomes an essential real-world factor for usage that cannot be overlooked for very large datasets. We compare the per-epoch training time of each method. Models were trained for one epoch on a standalone server, as reported in Section 4.3, to ensure the same computing resource was utilized for evaluating the training time for all models except INCV; the training time for the INCV model exceeded three days, which is notably longer than that of other methods. For fair comparison, we also consider the time for pre-training steps, including the warm-up epochs for DivideMix and the training time for the auto-encoder in our method (which is ~45 minutes in practice). We can observe from Fig. 5 that the additional calculation of inter-cluster and intra-cluster distances is insignificant. Since Co-Teaching and DivideMix train two models simultaneously, the training time is nearly double that of our method.

## 5.5 Explore CMC latent space

To investigate the effect of the CMC loss on the latent space, we explore the cohesiveness of its learned latent space. To proceed, we randomly selected 100 samples from the Institution B waveform dataset and retrieved the top-50 nearest neighbors in the latent space from the training set for each of the selected samples. The representation from the last convolutional layer was used to calculate the Euclidean distance to find the nearest neighbors. We manually annotate the selected neighbors, and we define "true label" if the label of related alarm type (AF or Non-AF) is the same as manual annotation. For each sample, we calculate purity (i.e., the percentage of true labels) in its neighborhood. Furthermore, we calculate the counter-class ratio (CCR), which represents the ratio of true-label samples that were incorrectly predicted among the 50 nearest neighbors of the test sample. By analyzing the latent space in this way, we gain a better understanding of the influence of the CMC loss on the cohesiveness of the latent space. A higher purity and lower CCR would indicate that the CMC loss induces the latent space to be more resilient towards label noise, leading to more robust AF detection.

As shown in Figure 7a, we observe that the neighbors obtained from CMC had a higher purity, indicating higher concentration of samples with correct labels, compared to those obtained from CE. In Figure 7b, we examined whether the neighbors obtained from CMC and CE belonged to the same class as the selected sample. We found that the neighbors obtained from CMC had significantly fewer counter-class neighbors compared to CE, indicating better separation between classes leading to more robust classifications.

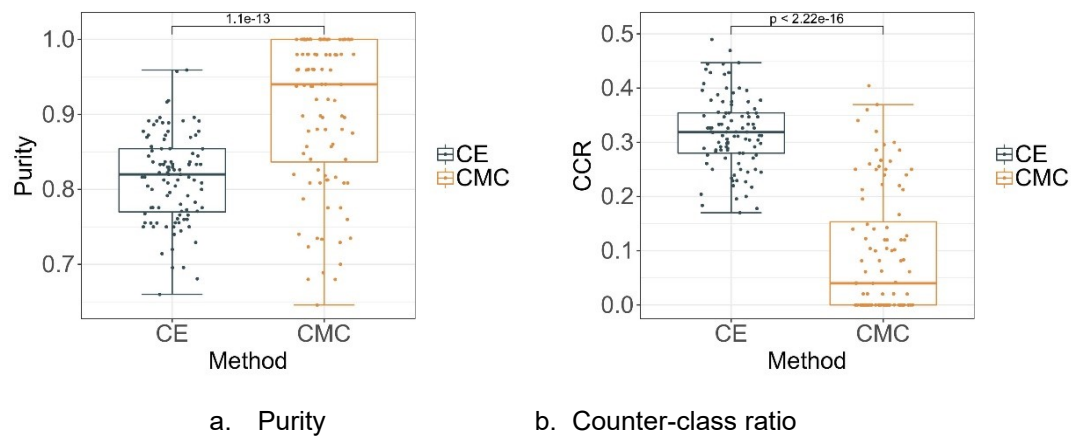

a. Purity      b. Counter-class ratio

Figure 7. Purity (a) and counter-class ratio (b) of the 50 nearest neighbors obtained from the training data. This figure provides a visual comparison of the performance of the two methods in identifying relevant neighbors from the training data for each selected sample.

# 6   Discussion

In this study, we used cardiac arrhythmia alarms generated from bedside patient monitors to build by far the largest dataset to train PPG-based AF detection models. The dataset contains over 8 million 30-second PPG segments obtained from over 24,000 hospitalized patients. Even though the AF labels in this dataset are noisy due to the existence of false cardiac arrhythmia alarms, our results on three external test datasets show that using this large dataset to train a CNN-based AF detector is still beneficial even without explicitly relabeling the data.

We also proposed a novel approach to improve learning from this large dataset with noisy labels. Our approach uses an unsupervised learning approach (autoencoder + clustering) to cluster training samples and uses the cluster membership of each sample to regularize the latent representation of a supervised network for PPG signals learned with noisy labels. This novel approach achieves superior or comparable results when evaluated against five different state-of-the-art robust learning algorithms for noisy labels, while at the same time maintaining a computational efficiency advantage. Specifically, it trains in half the time of its best competitor, DivideMix. Through experiments, we also demonstrated the effectiveness of combining our large scale auto-labeled dataset and our proposed robust learning approach for the important PPG AF detection task.

## 6.1   The value of large dataset with noisy labels

Large scale data with clean labels is difficult to obtain because extensive annotation is costly, time-consuming, and difficult to scale; also, manual annotation at a large scale does not guarantee error-free results due to multiple sources of labeling errors, particularly annotator fatigue. In this study, we build a large-scale PPG dataset leveraging imprecise AF alarms. From Fig. 3, we conclude that, even with inaccurate labels, the same algorithm trained on a large-scale dataset outperforms those trained on a clean labeled dataset that is much smaller. This result further motivated us to pursue novel approaches to explicitly address the issue of noisy labels. We acknowledge that the severity of the noisy label issue in our dataset remains unknown because we did not annotate all AF and PVC alarms used to sample PPGs. However, the distribution of incorrect labels in such a dataset represents a real-world situation – this is in contrast to the simulated noisy labels used in many studies of robust learning. Therefore, the benefits we demonstrate in the present study are likely obtainable by other researchers with access to similar real-world datasets.

## 6.2  Alternative approaches to auto-labeling PPG

To our knowledge, we are the first team to auto-label physiological signals with an *a priori* understanding that the generated labels may not be perfect. This is well motivated by prior work that used simulated label noise and demonstrated that deep neural networks are robust to label errors [51]. However, the strategy used in this study to auto-label the signal can still be improved as we used only alarm information. More modalities from EHR, such as patient diagnosis, demographic features, and charted events could be used to further filter false alarms and hence improve the accuracy of auto-generated labels. Moreover, the computerized algorithms behind alarm generation are based on ECG signals, so another future direction will be to use state-of-the-art ECG-AF algorithms to label PPG data.

## 6.3  CMC for label noise alleviation

The main finding of our study is that cluster membership consistency can mitigate the impact of label errors in a real-world dataset. Our proposed method and DivideMix outperform other methods. This is most likely due to the inclusion of unsupervised information in both methods. The cluster membership information used in our CMC approach is derived from an unsupervised learning process prior to conducting supervised learning with noisy labels. DivideMix treats the label noise issue as a semi-supervised problem, dividing the training data into clean and noisy subsets, and re-labeling the noisy subsets by "co-guessing" from two pre-trained networks to ensure consistency. Our investigation of the latent space further explains the effects of incorporating cluster membership information. As shown in Fig. 7, the incorporation of CMC loss is able to place the representation of each input PPG sample into a neighborhood where more training samples carry true labels (i.e., the auto-generated label is the correct label and belongs to the same class, AF or non-AF, as the input PPG sample). Such behavior would logically translate into better classification accuracy by a classifier using learned PPG representations. ELR and SOP are two top performers from a recent competition of state-of-the-art robust learning algorithms in computer vision. However, they exhibited poorer performance than our method and DivideMix. Potential reasons can be summarized as follows: a) Real-world noise. ELR and SOP were shown to be superior based on experiments that used simulated label errors. SOP was developed based on the premise that the model is over-parametrized. However, given the scale of our dataset and the model complexity, it may not satisfy the optimal conditions required for the success of the SOP algorithm. b) Poor data quality. Images in the datasets used in these two robust learning studies presumably have much lower signal-to-noise ratio than PPG signals. ELR relies on loss calculated during training to gauge the early learning stage, which could be unreliable due to noise in the data itself (poor signal quality): poor signal quality may cause loss to increase regardless of label correctness. The key assumption of ELR is that early-stopping can prevent the model from overfitting caused by label noise. However, input data quality as a factor has not been investigated, and our results show that early-stopping cannot be trusted when poor data quality is involved. c) We use an autoencoder as the preliminary representation step for clustering. Auto-encoders aim to learn an informative compressed representation that can be recovered back to the input space, which would remove some artifacts in the signal during the recovery (decoder) process. In our study, the learned representation for each signal is robust to poor signal quality to a certain extent, which is a plausible reason for our good performance on poor-quality data.

## 6.4    Scalability and Flexibility

In Figure 6, it is evident that the proposed CMC approach incurs only two extra minutes of training time compared to the conventional Cross-Entropy (CE) approach. In contrast, to achieve comparable performance, DivideMix requires training two models concurrently with additional pre-trained warm-up epochs. Therefore, CMC demonstrates superior scalability, which makes it a promising solution to handle label noise for large-scale datasets. Furthermore, in comparison to Co-Teaching and DivideMix, which require an additional step to adapt to new tasks, our proposed approach is more flexible and can be easily integrated with different model architectures and tasks, thus making it a plug-and-play option.

## 6.5 The choice of cluster number

We are aware there are techniques to find an optimal cluster number, such as Silhouette analysis [52] and Calinski-Harabasz Index [53]. Due to limited computation resources, we did not comprehensively experiment with the choice of cluster number. We chose 6 as the cluster number because we used three types of alarms, AF, PVC and NSR, and signals from each alarm type can be of good quality or poor quality, which gives us 6 categories total. However, results as presented in both Table 4 and Fig. 4 did not show a significant difference among different cluster numbers. Also, the goal of this study is to propose a generic and applicable strategy for label noise, whereas the choice of cluster number is a hyperparameter and is task-specific. A comprehensive investigation of finding the optimal cluster number is important for future work.

## 6.6 Limitations

Resnet-34 was used as the base model to compare different robust learning approaches, although our first experiment shows that a deeper network can perform better. Therefore, it remains to investigate whether CMC and other robust learning approaches can still significantly improve with deeper networks. Although our model excels in detecting AF in scenarios using wearable PPG in hospital settings (i.e., collected under specially defined conditions), more work is needed to test its performance on data obtained from wearables in free-living settings where the clinical significance of using PPG to screen AF mostly lies.

# 7 Conclusion

This study used cardiac arrhythmia alarms from bedside patient monitors to construct the largest current dataset for training PPG-based AF detection models, comprising over 8 million 30-second PPG segments from more than 24,000 hospitalized patients. Despite the presence of noisy AF labels caused by false alarms, our results on three external test datasets demonstrated the advantages of training a CNN-based AF detector using this extensive dataset, without the need for explicit data relabeling. To further enhance learning from this dataset with noisy labels, we proposed a novel CMC loss. Through comprehensive experiments, we validated the effectiveness of combining our large-scale auto-labeled dataset and the proposed robust learning approach for PPG AF detection. These findings contribute to the advancement of the field and present opportunities for further improvement in the accuracy and efficiency of AF detection models in clinical settings.

## Acknowledgment

This work was partially supported by NIH grant awards including R01HL166233, K23HL125251, and R01HL155711. We also highly appreciate the annotation effort from Masunaga, Colette, Lee Jenny and Dr. Zhe Zhang.

## Data Availability

Requests to access private datasets used in this study should be directed to the corresponding author. Such requests will be evaluated according to the existing data use agreement between institutions and may require escalation to appropriate institutional data governance committee(s).

## Code Availability

All code of this work can be accessed at https://github.com/chengding0713/Cluster-membership-consistency. As an alternative to researchers who are just interested in testing models reported in this paper, we hosted all the trained models in this study through a web application ModelMeetsData (M$^2$D) at https://nursingdatascience.emory.edu/modelmeetsdata. Using M$^2$D, users can conduct inference (detecting AF) using these models by passing their own data through M$^2$D. In the default mode of M$^2$D, data supplied by users will not be retained by M$^2$D after model inference is completed.

## Author contribution

C.D., Z.C., R.X., C.R., G.C. and X.H. designed the study and supervise all the process. R.J.L., F.N., A.S. and D. D. provided the clinical insight. C.D. and Z.C. did the data analysis and deep model training. C.D., R.X. and Z.C. wrote the first draft of the manuscript. All authors have read and agreed to the published version of the manuscript.

## Conflict of interest

All authors claim that they have no conflict of interest.